\documentclass[a4paper,fleqn]{cas-sc}

\usepackage[authoryear,longnamesfirst]{natbib}
\usepackage{amsmath,amssymb}
\usepackage{booktabs}
\usepackage{graphicx}
\usepackage{xurl}   

\newcommand{\CodeRepoURL}{\url{https://github.com/tkEzaki/llm-freight-market}}

\makeatletter
\renewenvironment{figure*}[1][!htb]{\@float{figure}[#1]}{\end@float}
\makeatother

\begin{document}
\let\WriteBookmarks\relax

\shorttitle{When Shippers Become Algorithms}
\shortauthors{T.~Ezaki et~al.}

\title[mode = title]{When Shippers Become Algorithms: Candidate Exposure,
Information Design, and the Concentration of LLM-Mediated Freight Markets}

\author[1]{Takahiro Ezaki}[orcid=0000-0003-4175-3028]
\cormark[1]
\ead{tkezaki@g.ecc.u-tokyo.ac.jp}
\credit{Conceptualization, Methodology, Software, Formal analysis,
Writing -- original draft}

\author[1]{Naoto Imura}[orcid=0000-0002-1428-7407]
\credit{Supervision, Writing -- review \& editing}

\author[1,2]{Katsuhiro Nishinari}[orcid=0000-0003-3069-1757]
\credit{Supervision, Writing -- review \& editing}

\affiliation[1]{organization={Research Center for Advanced Science and
    Technology, The University of Tokyo},
    addressline={4-6-1 Komaba, Meguro-ku},
    city={Tokyo},
    postcode={153-8904},
    country={Japan}}

\affiliation[2]{organization={Department of Aeronautics and Astronautics,
    School of Engineering, The University of Tokyo},
    addressline={7-3-1 Hongo, Bunkyo-ku},
    city={Tokyo},
    postcode={113-8656},
    country={Japan}}

\cortext[1]{Corresponding author.}

\begin{abstract}
Shippers are beginning to delegate carrier selection to large language
model (LLM) agents. We ask what such delegation does to a freight
matching market, and which platform design choices contain it. We
carried out agent-based simulations in which fifty shipper agents,
built on commercial LLMs from OpenAI (GPT), Anthropic (Claude), and
Google (Gemini), procure truckload capacity for thirty days. The market
implements the rules of digital freight matching: each load is offered
down the shipper's ranked list of carriers (waterfall tendering),
carriers have daily capacity limits, spot prices respond to congestion,
and carrier ratings accumulate with transactions. We found three risks
and one remedy that works.
Agents converged at once: for a fixed sampled carrier population, the
same carrier was the modal first choice of every model on day one,
attracting up to $76\%$ of requests. Because
each agent picks from its own randomly drawn list of displayed
candidates, the platform controls
how many options each shipper sees; concentration rose steeply once
lists
exceeded about ten carriers, with the onset differing across models.
Which carriers ended up dominant varied widely from one sampled
market to another, and displaying true quality instead of estimated
ratings changed neither the level nor this variability (by design,
quality affects only what agents see, never delivery outcomes).
Against these
risks, disclosing each
carrier's remaining daily capacity cut concentration by a third and
doubled shipper surplus, while vendor diversification, list-order
randomization, and popularity display showed no clearly detectable
effect. Platform
information design, ahead of model choice or model regulation, is the
lever that works.
\end{abstract}

\begin{highlights}
\item First simulation of freight procurement by populations of real LLM
agents with endogenous capacity, pricing, and ratings.
\item Choices concentrate on one carrier on day one; capacity limits and
congestion pricing then spread demand and give unrated entrants work.
\item Concentration rises steeply once each shipper's randomly drawn
candidate list exceeds about ten carriers; the onset depends on which
LLM the
shippers use.
\item Final market structure varies widely across sampled markets
under every trust signal tested; displaying true quality does not
reduce this variability.
\item Disclosing carriers' remaining capacity lowers concentration and
raises shipper surplus; diversification and list shuffling do not.
\end{highlights}


\maketitle

\section{Introduction}
\label{sec:intro}

Freight procurement is becoming a task that firms delegate to machines.
Large language models (LLMs) already sit inside supply chain workflows,
where they analyze contracts, screen suppliers, and forecast demand
\citep{wang2025llm,song2026llm,nie2025joint}, and prototype multi-agent
systems automate vendor scoring and selection end to end
\citep{joshi2026vendor}; the first commercial freight-procurement
agents that automate carrier sourcing were announced in 2026
\citep{project44agent}. Digital freight matching (DFM) platforms such as
Uber Freight and Full Truck Alliance have used algorithms to price loads
and recommend carriers for a decade. The step now underway is different
in kind. Shippers are handing the choice of carrier itself to autonomous
agents that read a list of candidates and pick one. When thousands of
shippers choose through a handful of foundation models, the matching
process inherits whatever preferences those models share.

They share a great deal. On the first simulated day of the experiments
reported below, with the carrier population held fixed, GPT agents
sent $34$ to $38$ of $50$ requests to the
same carrier out of twenty, and they did so under each of the three
information conditions we study (a list showing quoted prices together
with true reliability scores, with frozen ratings, or with live customer
ratings); Claude and Gemini agents made that carrier their most frequent
first choice as well. Which carrier plays this role depends on the
sampled population, not on the model (Section~\ref{sec:res-dayone}). This is \emph{algorithmic monoculture}
\citep{kleinberg2021,peng2024} materialized in a market. Independently
deployed systems have been shown to hand the same individuals the same
outcomes \citep{bommasani2022}, and LLM
populations converge on collective conventions \citep{ashery2025}. In a
matching market with capacity constraints, this creates an immediate
operational problem: a carrier chosen by everyone cannot serve
everyone's demand, so most of that demand has to go somewhere else.

Existing monoculture results are static. They study one-shot allocation
and abstract from the feedback that markets supply. Carriers raise prices
where demand piles up; capacity limits push displaced shippers onto
second choices; ratings accumulate for whoever gets chosen and stay
absent for whoever does not. Research on human markets describes how
such feedback can go either way. Herding models show that observational
and reputational signals can lock markets into inefficient,
path-dependent outcomes \citep{banerjee1992,bikhchandani1992,smith2000},
and the MusicLab experiment showed that social signals raise both
inequality and unpredictability of success
\citep{salganik2006}, with rating-system field experiments finding
related herding effects \citep{muchnik2013}. Studies of platform reputation show
that ratings inform but also inflate and lag
\citep{tadelis2016,filippas2022}, and that entrants without a track
record face cold-start exclusion \citep{pallais2014}. No study yet
follows LLM choosers, whose biases are documented and correlated, through
a market whose prices, capacities, and ratings respond to their choices.
Work on LLM agents in markets has shown that pricing agents collude
without being instructed to \citep{fish2024} and has established LLMs as
usable simulated economic agents \citep{filippas2024homo}. The matching side,
where monoculture should matter most, remains open.

Freight is the right industry for this question, for two reasons. First,
truckload procurement already has well-defined rules that make the
problem concrete and measurable. Shippers keep ranked carrier lists
(routing guides) and offer each load down the list (waterfall
tendering); carriers reject loads when they run out of capacity; and
spot prices carry a persistent premium over contract rates
\citep{acocella2023,acocella2020}. Second, the tension our agents
exhibit, loyalty to incumbent carriers versus opportunistic spot search,
is the central empirical fact of this industry for human shippers
\citep{harris2025ltr,harris2025spot}. Outcomes in this setting carry
managerial meaning: service rates, freight rates, shipper surplus, and
carrier revenue concentration, rather than abstract network statistics.
Recent work quantifies the information value of freight platforms
\citep{wu2024}. We extend that agenda to the case where the
platform's users are themselves algorithms.

We build a simulation platform in which real LLM agents procure freight
inside a market with endogenous economics. Fifty shipper agents, powered
by commercial models from OpenAI, Anthropic, and Google, choose among
twenty carriers every day for thirty days. The market implements
waterfall tendering over ranked preferences (each load goes to the
agent's first choice if that carrier has capacity left, otherwise to the
second, then the third), a fixed daily capacity limit for each carrier,
spot prices that adjust daily to excess-demand pressure, and a rating
system in which a carrier's displayed score aggregates the reviews of
its completed shipments. A carrier that no one chooses accumulates no
reviews. We manipulate three levers, each under the platform's control:
the \emph{exposure} $L$, the number of candidate carriers an agent can
choose from ($L \in \{3,5,10,15,20\}$); the \emph{information structure}
of the trust signal (the true reliability disclosed directly, a rating
frozen at its initial value so that it never updates, or the live
customer rating); and \emph{disclosure interventions}, including
real-time capacity visibility, list-order randomization, and popularity
display, plus a vendor-mix condition that splits shippers across the
three LLMs. We report 226 experimental cells, five to ten independent runs per
condition and about 190{,}000 individual carrier choices, and we log
every prompt and response for audit.

We found four things. First, the concentration onto a single favorite
forms on day one under every information structure; the rules of the
market then determine how much of it survives. Two mechanisms work
against it. The crowded carrier's price spikes, and price-sensitive
agents respond by trying cheap unrated entrants, whose ratings converge
to their true quality within thirty days. The \emph{cold-start trap}
that reputation theory predicts (entrants without ratings never get
chosen, so they never earn ratings; \citealp{pallais2014}) did not
arise, because high prices at the crowded carrier keep pushing trials
toward newcomers. Second, concentration rises sharply with exposure
for two of the three vendors, with most of the rise between $L=10$
and $L=15$, while the third (Anthropic's Claude)
stays flat at every exposure. A deterministic scoring baseline, by
contrast, concentrates from small $L$ onward; a decomposition
attributes the shape of the rise largely to candidate-set overlap and
the vendor contrast to the models themselves. Third, the surviving
market structure varies widely across independently sampled markets,
and most of that variation traces to the drawn market rather than to
the trust signal: replacing estimated ratings with disclosed true
reliability changed neither the level nor the spread of final
concentration. Fourth, of the four interventions, only
capacity disclosure had a clearly detectable effect. Displaying
remaining daily capacity next to
each candidate cut concentration by a third and doubled shipper
surplus; vendor diversification, list shuffling, and popularity display
showed no clearly detectable effect. Congestion management, which
\citet{arnosti2021} implement by limiting applications and we
implement through disclosure, is the effective lever in AI-mediated
matching as well.

The paper makes three contributions. We provide, to our knowledge, the
first freight-specific market-level testbed that embeds populations of
real LLM agents in
a matching market combining waterfall tendering, binding capacity,
congestion pricing, and endogenous ratings, with
a decision-level audit trail; the platform extends to other two-sided
logistics questions. We recast algorithmic monoculture as a contest
between two feedback loops, rating accumulation that entrenches the
early favorite and congestion pricing that works against it, which
connects the monoculture literature \citep{kleinberg2021,peng2024} to
herding and market design \citep{banerjee1992,roth2008} and explains why
static analyses overstate lock-in. For practice, we give DFM platforms
design guidance for the arrival of AI shippers: keep the exposure knob
below the transition region, disclose capacity, and expect little
from model
diversity or list randomization.

The remainder of the paper proceeds as follows.
Section~\ref{sec:related} positions our study in the literature.
Section~\ref{sec:model} describes the market institutions and the LLM
agent architecture. Section~\ref{sec:design} details the experimental
design. Section~\ref{sec:results} reports results, and
Section~\ref{sec:discussion} draws managerial and regulatory
implications, discusses limitations, and outlines extensions.

\section{Related work}
\label{sec:related}

\subsection{Algorithmic monoculture}
\citet{kleinberg2021} show that when many decision makers adopt the same
algorithm, welfare can fall even if that algorithm beats every
alternative each firm holds, because errors correlate across adopters.
\citet{peng2024} carry the argument into matching markets, where
correlated evaluations distort both match quality and its distribution
across participants. \citet{bommasani2022} document the empirical
signature, outcome homogenization: the same individuals receive the
same decision from independently deployed models.
\citet{anwar2024} find homogenization of consumption in long-run
recommender simulations, and \citet{ashery2025} show that populations
of interacting LLMs converge on shared conventions without any central
coordination. \citet{ballestero2026strategic} add experimental
evidence on the strategic side: in coordination games, LLM subjects
sustain their baseline similarity even when payoffs reward
divergence, where human subjects differentiate. These studies hold
the environment fixed while the
algorithms decide. We let the environment answer back. Prices,
capacities, and ratings in our market respond to what the agents do,
and the imprint of monoculture then splits in two: the day-one
consensus is as extreme as the static literature would predict, while
its persistence is a property of the institutions, not of the model
population.

\subsection{Herding, social influence, and reputation}
Sequential-choice models show how public signals crowd out private
information and lock markets onto arbitrary outcomes
\citep{banerjee1992,bikhchandani1992,smith2000}. The MusicLab
experiment gave the mechanism its sharpest empirical form: social
signals raised the inequality and, above all, the unpredictability of
success \citep{salganik2006}, and follow-up field experiments found the
ratings version of the same asymmetry \citep{muchnik2013}. The
platform-reputation literature documents both the value and the
pathologies of feedback systems: reputation shifts sales and survival
\citep{cabral2010}, ratings inflate and censor \citep{nosko2015,
filippas2022,tadelis2016}, and workers without a track record face
measurable cold-start exclusion \citep{pallais2014}. Our agents differ
from the decision makers in herding models in a way that matters: LLM
shippers share a prior rather than observe one another, so the cascade
is complete at $t=0$. What our experiments add is the market's reply.
The unpredictability of outcomes that \citet{salganik2006} induced
with social signals appears in our market as a baseline property of
the feedback institutions, present even when true quality is
displayed, and the cold-start trap does
not close on entrants because tender rejection, not subsidy, supplies
them with trials.

\subsection{Congestion and information design in matching platforms}
\citet{roth2008} lists congestion among the three failures a
marketplace must manage. \citet{arnosti2021} show that application
congestion destroys welfare in matching platforms and that limiting or
redirecting applications restores it; \citet{guo2023} analyze queue
disclosure as a signaling device; \citet{kamenica2011} give the
general framework in which a designer chooses what information to
release. In freight, \citet{wu2024} quantify the information value of
a logistics platform, and \citet{liu2025swing} design binding
multi-period contracts to hedge the same friction our model generates
endogenously, tender rejection under tight capacity. Our capacity
disclosure intervention belongs to this family: it is congestion
management executed inside the prompt, and it addresses with information
the same friction that \citet{liu2025swing} address with contracts.

\subsection{LLM agents in markets}
\citet{filippas2024homo} establish LLMs as simulated economic agents;
\citet{park2023} demonstrate believable multi-agent populations.
On the market side, \citet{fish2024} show that LLM pricing agents
reach supracompetitive prices without being instructed to collude, and
a fast-growing set of simulation studies examines governance,
information asymmetries, and
exploitation in LLM agent markets \citep{syrnikov2026,lei2026,erlei2026}.
Application work has moved procurement itself onto LLM rails, from
supplier evaluation frameworks \citep{joshi2026vendor} and
domain-specialized supply chain models \citep{wang2025llm} to reviews
of LLM adoption in supply chain management \citep{song2026llm},
LLM-based demand prediction \citep{nie2025joint}, and automated
design of routing optimizers \citep{wang2026routing}. Two testbeds
come closest to our aim. \citet{liu2026agentmarkets} build a
programmable marketplace in which tool-using agents post jobs,
negotiate, and accumulate reputation, and find that well-meant
institutional interventions can degrade rather than improve market
performance; \citet{long2026reliability} replace every player of a
serial supply chain with generative agents and identify run-to-run
decision instability, which they call agent bullwhip, as the central
reliability risk. Neither market clears under binding capacity with
congestion pricing and endogenous ratings, and neither asks the
concentration question. The matching side, where correlated
preferences meet
capacity constraints, has lacked a testbed with endogenous economics.
The simulation platform of this study provides one.

\section{Market model and agent architecture}
\label{sec:model}

\begin{figure*}[!htb]
\centering
\includegraphics[width=\textwidth]{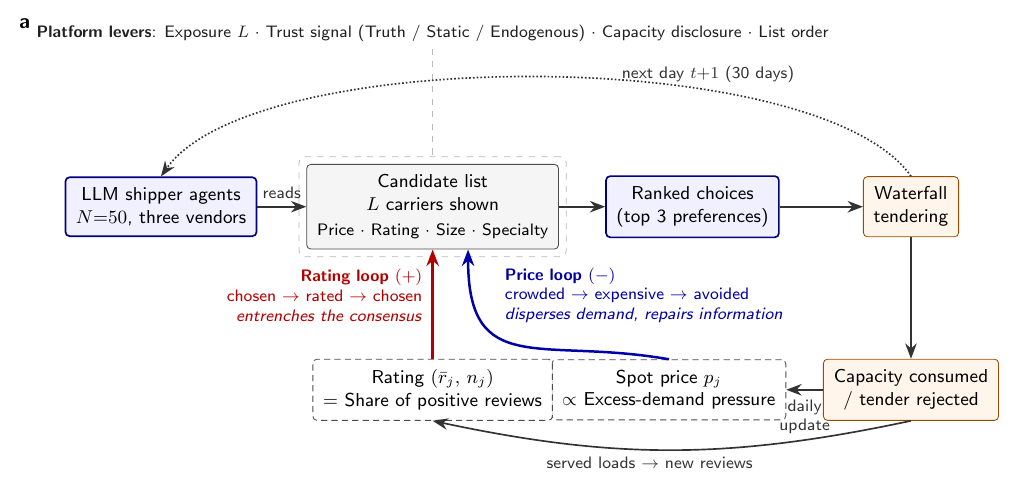}\\[2mm]
\includegraphics[width=\textwidth]{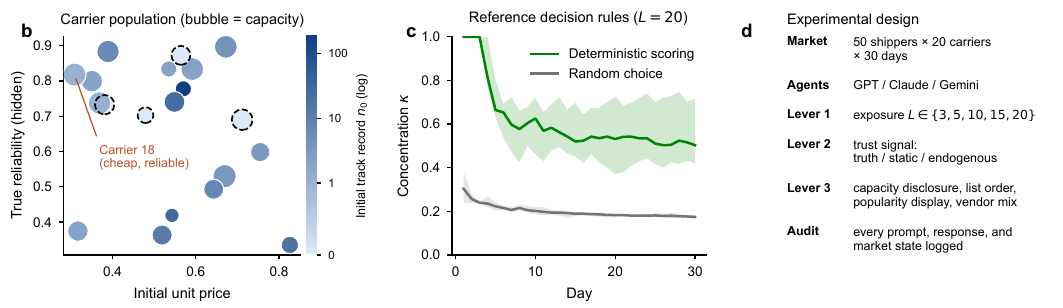}
\caption{\textbf{A freight matching market with two feedback loops.}
(a)~Daily loop. Shipper agents read a candidate list, return ranked
choices, and the platform tenders each load down the list under
capacity constraints. Ratings accumulate for chosen carriers (red loop)
while congestion pricing penalizes crowded carriers (blue loop). The
dashed box marks the levers we manipulate.
(b)~One sampled carrier population (seed 0). Bubble size gives daily
tonnage capacity; color gives the initial number of rated shipments
$n_0$ (log scale); dashed black circles mark carriers that start with
no ratings.
Carrier 18 combines a low price with high reliability.
(c)~Reference decision rules at $L=20$ (mean and seed range): the
deterministic common-score rule concentrates strongly, random choice
gives the floor.
(d)~Design summary; Table~\ref{tab:design} gives the full
condition-by-run accounting.}
\label{fig:model}
\end{figure*}

Figure~\ref{fig:model}a summarizes the simulated market. Fifty shipper
agents, each controlled either by a commercial LLM or by one of four
reference algorithms described below, procure truckload service from
twenty carriers over thirty days. Each day every shipper issues one load
with a random origin--destination pair, a dimensionless distance index
drawn uniformly from
10 to 100, and a weight drawn from $\{5,10,15,20,30\}$ tons; the payment
for a served load equals the carrier's unit price times distance times
tons, so all monetary quantities are expressed per ton-distance. The institutions follow truckload practice \citep{acocella2023}:
the shipper agent returns a ranked list of three carriers, the platform
tenders the load down that list, and the first carrier with remaining
daily capacity serves it. If all three ranked carriers are full, the
load goes unserved and the shipper incurs a delay cost of $0.3$, in
the same per-ton-distance units as prices. Within a day, loads are
processed in a random order drawn each day, which also arbitrates who
obtains scarce capacity. This waterfall
corresponds to the routing-guide mechanics that shippers use in practice
\citep{acocella2020}. We cap the waterfall at three tenders because each
additional tender costs the shipper coordination time and delay in
practice, and routing guides rarely go deeper before a load is
re-brokered; letting agents rank the full displayed list is a
straightforward extension we leave for future work.

\paragraph{Carriers.}
Each carrier $j$ has four static attributes drawn independently and
uniformly from $[0.3, 0.9]$: an initial unit price, a true reliability
$\rho_j$, a capacity weight, and a route specialty
(Fig.~\ref{fig:model}b). Daily tonnage capacity is allocated in
proportion to the capacity weight and calibrated so that aggregate
capacity equals $1.3$ times expected daily demand; individual carriers
handle 27 to 70 tons per day. One carrier per sampled population has
less daily capacity than the largest 30-ton load; it simply rejects
such loads in the waterfall. Unit cost equals $0.6$ times the initial
price, which fixes each carrier's margin.

\paragraph{Prices.}
Carriers reprice daily in response to excess-demand pressure:
\begin{align}
p_{j,t+1} &= \min\!\big( \max( \tilde{p}_{j,t},\; 1.05\,c_j ),\; 1.5 \big),
\label{eq:price}\\
\tilde{p}_{j,t} &= p_{j,t}\,[\,1 + \alpha\,(\pi_{j,t} - 1)\,],
\nonumber\\
\pi_{j,t} &= \frac{q^{\mathrm{srv}}_{j,t} + q^{\mathrm{rej}}_{j,t}}{Q_j},
\nonumber
\end{align}
where $p_{j,t}$ is carrier $j$'s unit price on day $t$, $c_j$ its unit
cost, $Q_j$ its daily tonnage capacity, $q^{\mathrm{srv}}_{j,t}$ and
$q^{\mathrm{rej}}_{j,t}$ the tons it served and rejected that day, and
$\alpha = 0.10$. In words: price moves by $10\%$ of the gap between
demand pressure $\pi_{j,t}$ and one, bounded below by $1.05$ times unit
cost and above by an index of $1.5$. A carrier that fills its book
exactly holds its price; a carrier that turns demand away raises it.
Because a rejected load continues down the routing guide, its tonnage
can enter the pressure of more than one carrier on the same day; this
is intended, since each rejection is a demand signal that carrier
actually turned away, rather than double counting.

\paragraph{Ratings.}
The true reliability $\rho_j$ stays hidden. After each completed
shipment the shipper leaves a review: a positive review with probability
$\rho_j$, a neutral one otherwise. Real reviews are richer than this
binary rule; we adopt it as a deliberate simplification that makes the
displayed rating an unbiased estimate of $\rho_j$, so that any market
pathology we find cannot be blamed on biased reviews. The platform
displays the running mean $\bar r_j$ of these reviews together with the
count $n_j$ of reviewed shipments. Initial track records are set as
$n_0 = \max(0,\, \mathrm{round}(200 f_j) - 10)$, where the fitness
$f_j$ is a Pareto draw (shape $1.5$) rescaled to at most $0.95$, so
the distribution of $n_0$ is heavy-tailed: in the population of
Fig.~\ref{fig:model}b, a few incumbents start with up to 180
reviewed shipments and four carriers start with none, displaying ``no
customer ratings yet''. A carrier's initial displayed rating is the
mean of $n_0$ simulated reviews drawn with the same Bernoulli rule.
Ratings update only when a carrier serves a
load, so selection and information accumulate together. Reviews are
distinct from the \emph{failed} flag that appears in an agent's
transaction history, which marks loads that none of the three ranked
carriers could serve. One design choice deserves emphasis: reliability
affects outcomes only through the reviews it generates. Whether a load
is served depends on capacity alone, and payments do not depend on
$\rho_j$, so the trust signal is a pure coordination device rather
than a predictor of delivery failures. The information ablation of
Section~\ref{sec:res-information} therefore measures the signal's
coordination role; a channel from $\rho_j$ to physical service
failure is a deliberate omission we return to in the limitations.

\paragraph{Shipper agents.}
Each shipper has a fixed procurement priority (cost, reliability,
capacity, specialty, or balanced; ten shippers each) and is controlled
by a decision rule. In the LLM conditions, the rule is a commercial model that
reads a prompt and answers in JSON with three ranked carrier
identifiers. The prompt shows the day's load, a summary of the shipper's
last five shipments (unserved loads flagged as \emph{failed}), and one
line per candidate carrier: the quoted total price with its unit-price
breakdown, the trust signal (which we manipulate below), the daily
tonnage capacity, and the specialty score. \ref{app:prompts}
reproduces a complete prompt and a raw response. We use gpt-5.4-mini
(OpenAI), claude-haiku-4-5 (Anthropic), and gemini-3.5-flash (Google) at
temperature $0.7$. GPT is the default vendor;
Claude and Gemini replicate the exposure sweep, and one condition assigns the
three vendors to shippers in near-equal shares (17/17/16 of the
fifty). When fewer than twenty candidates are displayed
($L < 20$), the $L$ candidates are drawn uniformly at random from the
active carriers for each load; candidate order is randomized
per prompt unless stated otherwise.

In the reference conditions, the rule is one of four simple algorithms:
\emph{random}, a uniform draw from the displayed list; \emph{preferential
attachment} \citep{barabasi1999}, which picks carriers with probability proportional to their
discounted accumulated matches (the same $0.95$-per-day discounted
record used for $\kappa$); \emph{conflicting attachment}, a rule of our
construction named after \citet{leung2016}, in which accumulated
matches first attract and then repel; and a
\emph{common-score} rule that ranks all candidates by one fixed
weighting of the displayed attributes (all parameters in
\ref{app:baselines}). The common-score rule is
deterministic and identical across shippers, which makes it the closest
analogue of a monoculture of literally identical classifiers; it serves
as the main-text contrast together with the random floor
(Fig.~\ref{fig:model}c). The two attachment rules stay near the random
floor throughout and are reported in \ref{app:baselines}.

\paragraph{Outcome measures.}
Concentration $\kappa_t$ is computed on a running record of first
choices that discounts the past: each first choice adds one unit to
the chosen carrier's total, and at the end of every day all totals are
multiplied by $0.95$ (relationship weights below $0.05$ are dropped),
giving the record an effective memory of about twenty days.
$\kappa_t$ is the share of this discounted record held by the three
largest carriers; on day one it coincides with the day's first-choice
share, and ``final $\kappa$'' averages
days 26--30. A realized version of the same index, computed on the
loads carriers end up serving rather than on first choices, tracks the
allocation after tender rejections. We also use the
Herfindahl--Hirschman index (HHI) of daily first choices, the sum of
squared choice shares across carriers, to measure how spread out each
day's demand is. Welfare and service outcomes are the service rate, the
mean paid unit price, shipper surplus, and carrier revenue
concentration. A served load is worth $1.2$ per ton-distance to its
shipper, so shipper surplus is $1.2$ minus the paid unit price,
averaged over loads, with unserved loads entering at the delay cost
of $-0.3$. \ref{app:metrics} shows that the results are
unchanged under these alternative concentration indices.

\section{Experimental design}
\label{sec:design}

We manipulate three platform levers (Fig.~\ref{fig:model}d). The
\emph{exposure sweep} varies the number of displayed candidates
$L \in \{3,5,10,15,20\}$ for each vendor. The \emph{information
ablation} replaces the endogenous customer rating (the live score
that updates with each review) with
either the disclosed true reliability (\emph{truth}) or a rating frozen
at its initial value so that it never updates (\emph{static}), holding
everything else fixed; the comparison isolates what the rating dynamics
contribute beyond the informativeness of the signal. The
\emph{intervention comparison} starts from the default condition (GPT,
$L=20$, endogenous ratings) and adds one intervention at a time:
real-time capacity disclosure (remaining tons shown next to each
candidate, updated within the day), fixed list order, popularity display
(a discounted count of recent transactions shown as a popularity
score), and the vendor mix that assigns
GPT, Claude, and Gemini to shippers in near-equal shares.

Every condition receives five independent runs that redraw the carrier
population, the load stream, and the review outcomes, except the three
trust-signal conditions, which receive ten runs each: their result is a
statement about dispersion across runs, and variance estimates from
five runs would be too coarse to compare. For a given run index, the
three information conditions share the carrier population, the load
stream, and the candidate draws, so comparisons across them are
paired; all other condition families redraw everything, and variation
across their runs therefore mixes differences in the drawn market
with stochastic history (we return to this in
Section~\ref{sec:res-information}). Throughout, final $\kappa$ per run is
the unit of analysis: variance comparisons use Brown--Forsythe tests,
location
comparisons use exact two-sided Mann--Whitney tests (no ties
occurred), paired Wilcoxon tests are used where conditions share
seeds, and we report
Holm--Bonferroni-corrected $p$-values alongside the raw ones within
each family of comparisons. We report 226 cells
(Table~\ref{tab:design}) and about 190{,}000
individual LLM decisions. The simulator
logs every prompt, raw response, and market state transition; the
full audit logs are released with the code and data.

\begin{table}[!htb]
\centering
\footnotesize
\setlength{\tabcolsep}{3.5pt}
\begin{tabular}{lccc}
\toprule
Family & Conditions & Runs & Cells \\
\midrule
Exposure sweep & 3 vendors $\times$ 5 exposures & 5 & 75 \\
Reference rules & 4 rules $\times$ 5 exposures & 5 & 100 \\
Information ablation & truth, static ($L{=}20$) & 10 & 20 \\
\quad(endogenous) & GPT $L{=}20$ cells $+$ 5 runs & 10 & 5 \\
Interventions & capacity, order, popularity & 5 & 15 \\
Vendor mix & GPT/Claude/Gemini & 5 & 5 \\
Price counterfactual & $\alpha \in \{0.05, 0.02\}$ & 3 & 6 \\
\midrule
Total & & & 226 \\
\bottomrule
\end{tabular}
\caption{Experimental design. \emph{Runs} counts independent
repetitions per condition; each run redraws the carrier population,
the load stream, and the review outcomes. The endogenous information
condition reuses the five GPT $L=20$ sweep cells and adds five more
runs, so only the additional cells are counted in its row.}
\label{tab:design}
\end{table}

\section{Results}
\label{sec:results}

\subsection{Choices concentrate on day one; the market spreads them out}
\label{sec:res-dayone}

\begin{figure*}[!htb]
\centering
\includegraphics[width=\textwidth]{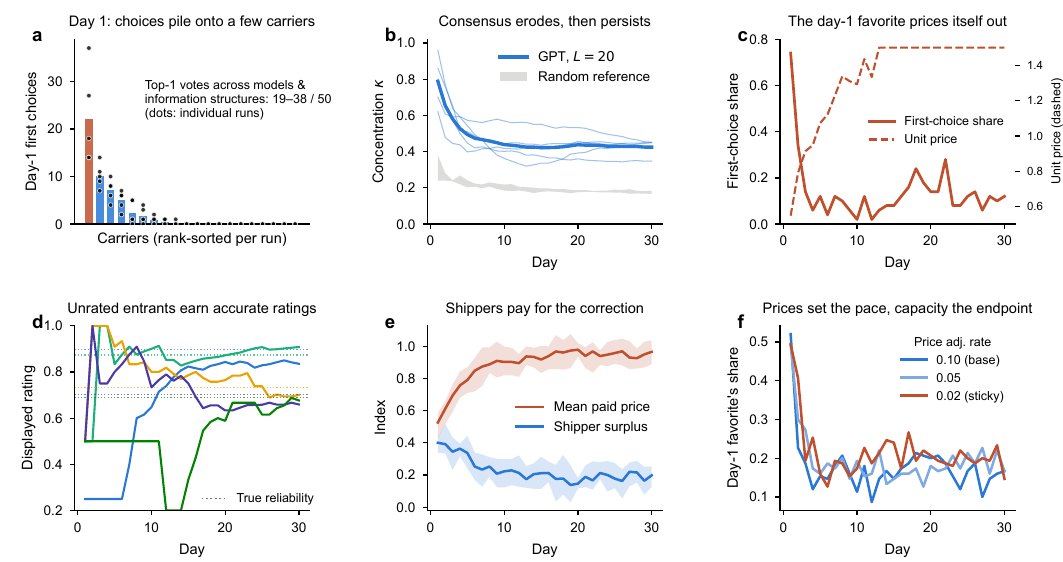}
\caption{\textbf{Choices concentrate on day one; capacity and prices
spread them out} (GPT, $L=20$, endogenous ratings).
(a)~Day-one first choices, carriers sorted by rank within each run:
bars give the mean across five runs, dots individual runs. The
top-ranked carrier draws 14--37 of 50 first choices per run. For a
fixed carrier population, the same carrier tops every vendor and
information condition (19--38 of 50).
(b)~Concentration $\kappa$ over 30 days: thin lines are runs, the
thick line the mean, gray the random-choice band.
(c)~The day-one favorite's first-choice share (solid) and unit price
(dashed) in the run of panel a: the price reaches the cap by day 13
and the share collapses.
(d)~Displayed ratings of the five initially unrated or underrated
carriers converge to their true reliabilities (dotted), same run.
(e)~Mean paid price and shipper surplus (mean $\pm$ range across runs).
(f)~Share of the day-one favorite under three price-adjustment rates
(means over three runs): stickier prices slow the decline without
changing its endpoint.}
\label{fig:dayone}
\end{figure*}

Figure~\ref{fig:dayone}a shows how first choices were distributed on
the first day of the default condition. The single most-requested
carrier drew 14 to 37 of 50 first choices depending on the run, and
the three most-requested drew 62 to 96\%; day-one $\kappa$ averaged
$0.79$, far above anything the market later sustained. Which carrier
plays the favorite depends on the drawn carrier population, not on
the decision maker: each run redraws the population and elevated a
different favorite, but for a fixed population every vendor and
information condition produced the same modal first choice. Under the
population of Fig.~\ref{fig:model}b, carrier 18,
the cheapest of the highly reliable incumbents, topped the day-one
tally in the default condition, under both other vendors, and under
both alternative information structures, with 19 to 38 of 50 first
choices.

This initial concentration then declined steadily over the first week
and settled at $\kappa = 0.42$ (range $0.35$--$0.45$ across runs),
roughly $2.4$ times the random-choice floor of $0.18$
(Fig.~\ref{fig:dayone}b). Figure~\ref{fig:dayone}c shows what happened
to the day-one favorite itself: its excess demand drove its unit price
to the cap within 13 days, and its first-choice share fell from $0.74$
to about $0.12$. Displaced and price-sensitive agents tried the cheap
carriers that started with no or almost no reviews, and those carriers'
displayed
scores converged to their true reliability within the month. For
example, the strongest of these entrants (true reliability $0.90$,
displayed as $0.25$ from four early reviews) served 81 loads and
ended at a displayed rating of $0.84$ (Fig.~\ref{fig:dayone}d). The
\emph{cold-start} exclusion that field evidence documents for human
markets \citep{pallais2014} did not appear here, because congestion
kept routing trials toward newcomers. Shippers paid for this partial
self-correction: the mean paid price rose from $0.53$ to $0.96$ and
shipper surplus halved (Fig.~\ref{fig:dayone}e).

A counterfactual that slows the price response separates the two market
forces behind the correction (Fig.~\ref{fig:dayone}f). At adjustment
rates of $0.05$ and $0.02$ instead of $0.10$, the favorite's decline
slowed (mean day-2 share $0.30$ and $0.41$ versus $0.23$) but arrived
at the same endpoint: by day 30 the day-one favorite held a share of
$0.15$--$0.17$ at every rate, and the initially unrated carriers still
served $160$ loads under the stickiest prices ($196$ at the base
rate). Tender rejection under the capacity constraint did the
dispersing and supplied the trials; congestion pricing set the pace.
The cold-start trap stayed open even when prices barely moved.

\subsection{Concentration rises sharply with exposure, and the rise
depends on the vendor}
\label{sec:res-exposure}

\begin{figure*}[!htb]
\centering
\includegraphics[width=\textwidth]{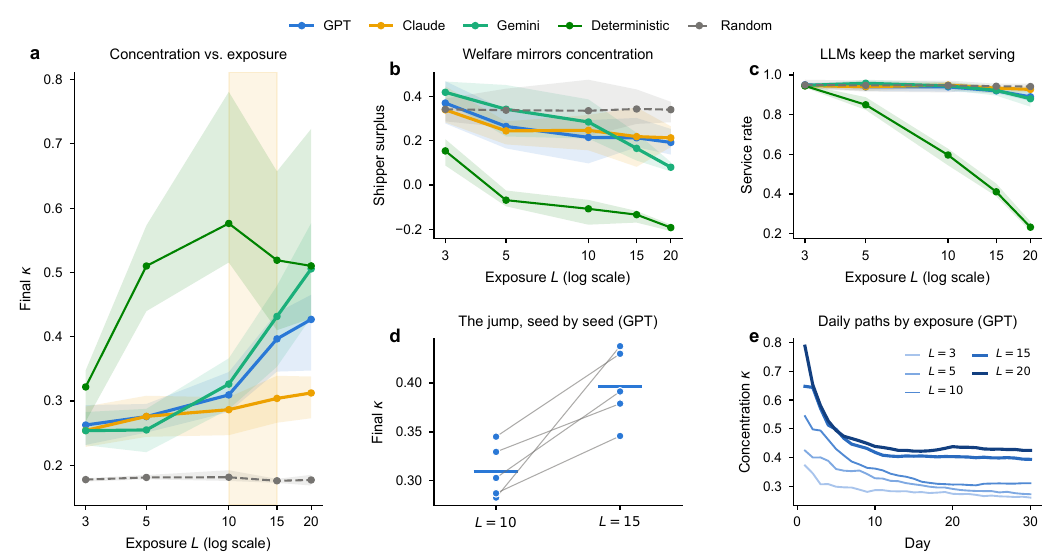}
\caption{\textbf{Concentration responds to exposure with a
vendor-dependent transition.}
(a)~Final $\kappa$ against exposure $L$ (mean $\pm$ range across runs;
bands are drawn where three or more runs are available; the shaded
vertical band marks the transition zone).
(b)~Shipper surplus and (c)~service rate against $L$.
(d)~Final $\kappa$ at $L=10$ versus $L=15$ for GPT; each dot is one
run, lines connect runs.
(e)~Daily concentration for GPT by exposure (means across runs).}
\label{fig:exposure}
\end{figure*}

Figure~\ref{fig:exposure}a traces final concentration against the
exposure $L$. For GPT the curve stayed flat at $\kappa = 0.26$--$0.31$
up to $L = 10$ and then climbed steeply, reaching $0.40$ at $L=15$ and
$0.43$ at $L=20$. Every run increased
from $L=10$ to $L=15$ (Fig.~\ref{fig:exposure}d). Gemini showed the
same shape with a higher ceiling ($0.51$ at $L=20$), and Claude never
left the $0.25$--$0.31$ band at any exposure. The deterministic
common-score rule, in contrast, concentrated from small $L$ onward
(the two attachment rules stayed near the random floor; see
\ref{app:baselines}).

Part of this rise is built into the display mechanics. The $L$
displayed candidates are a fresh uniform draw per load, so raising
$L$ does two things at once: it makes different shippers' lists
overlap more, and it raises the probability that any given carrier,
including the favorite, is on someone's screen ($L/20$). At small
$L$ the agents could not all pick the same carrier even if they
wanted to. To separate these mechanics from agent behavior, we ask
two questions of the data. First, do agents seize the favorite more
eagerly when lists are long? No: conditional on the run's day-one
favorite actually being displayed, the probability that an agent
picks it \emph{falls} from $0.57$ at $L=3$ to $0.14$ at $L=20$. The
rise in $\kappa$ therefore comes from the favorite appearing on more
screens, not from stronger attachment per encounter. Second, how
much of the rise appears with no change in agent behavior and no
market feedback at all? We freeze each run's day-one choice
frequencies and replay them, redrawing only the displayed subsets:
this static benchmark yields daily top-three shares of
$0.44/0.57/0.73/0.78/0.80$ at $L=3/5/10/15/20$. Candidate-set
overlap alone thus reproduces a rise of the same shape, concentrated
below $L=15$, though at a level far above what the interacting
market sustains. Together the two checks assign the roles: the
display mechanics produce the shape of the rise, the market's
capacity and price feedback compress its level, and the
LLM-specific content is the vendor contrast, since Claude faces
identical display mechanics at every $L$ and does
not rise. Whether model-level mechanisms such as long-context
retrieval degradation \citep{liu2024lost} contribute on top of the
mechanics would require designs that vary list length while holding
display probability fixed, which we leave to future work.

Wider exposure cost shippers money. Surplus fell with $L$ for all
three vendors while the service rate stayed near $0.9$
(Fig.~\ref{fig:exposure}b,c); the deterministic rule shows what a
market breakdown would look like instead, with service collapsing to
$0.23$ at $L=20$. The daily trajectories (Fig.~\ref{fig:exposure}e)
show where the transition operates: every exposure started
concentrated, but only at $L \ge 15$ did the day-one concentration
survive the month.

\subsection{Information quality had no detectable effect on the level
or the spread}
\label{sec:res-information}

The second lever is the trust signal itself. The default is the
\emph{endogenous} rating, the live customer score that updates as
reviews arrive. Holding the market fixed, we replaced it with either
the true reliability (\emph{truth}) or a rating frozen at its initial
value (\emph{static}), and asked whether the quality of the displayed
information changes what the market converges to.

\begin{figure*}[!htb]
\centering
\includegraphics[width=\textwidth]{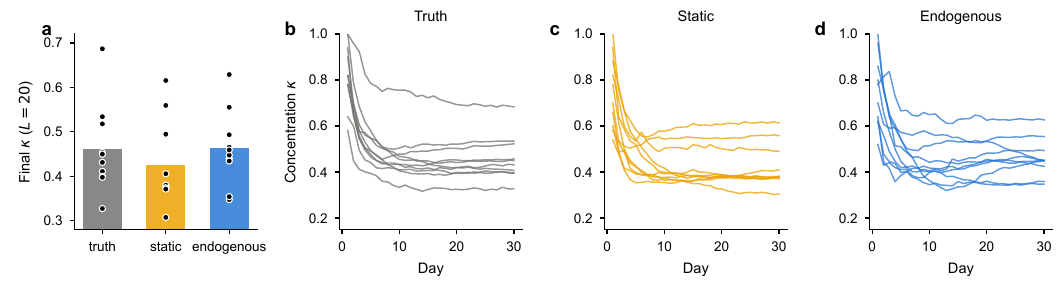}
\caption{\textbf{Final concentration varies widely across sampled
markets under
every trust-signal structure} (GPT, $L=20$).
(a)~Final $\kappa$ by trust-signal structure; bars give means across
runs, dots individual runs (ten runs per structure).
(b--d)~Daily trajectories per structure, one line per run: under
every structure the runs split into locked-in and dispersed
branches.}
\label{fig:information}
\end{figure*}

The answer is that no effect is detectable. Mean final
concentration barely moved across the three structures ($0.46$ under
truth, $0.43$ under static, $0.46$ under endogenous;
Fig.~\ref{fig:information}a; paired Wilcoxon $p \ge 0.19$, unpaired
Mann--Whitney $p \ge 0.21$, for
every pairwise comparison). Neither did the dispersion: the standard
deviation of final
$\kappa$ was $0.10$ under truth, $0.10$ under static, and $0.08$
under endogenous (Brown--Forsythe $p \ge 0.75$ for every pairwise
comparison; no pairwise difference in means exceeds $0.04$, and every
bootstrap $95\%$ confidence interval brackets zero, the widest being
$[-0.12, 0.05]$).
Under every structure, individual runs ranged from
dispersed ($\kappa = 0.31$--$0.35$) to locked onto an arbitrary
favorite ($\kappa = 0.62$--$0.69$), and the daily trajectories split
into branches that never rejoined (Fig.~\ref{fig:information}b--d).

What produces this wide dispersion? Because runs redraw the carrier
population and the load stream, variation across them mixes
differences in market fundamentals with stochastic history, and the
shared seeds across the three information conditions let us bound
the two parts. Final $\kappa$ is strongly correlated across
conditions run on the same population ($r = 0.93$ between truth and
endogenous, $0.49$--$0.68$ for the other pairs), and a variance
decomposition attributes $76\%$ of the total variation to which
market was drawn, with the remaining $24\%$ arising within a market
across conditions and stochastic histories. The dispersion is
therefore primarily a property of the sampled market: even with each
carrier's true reliability on display, which carriers the market
elevated depended on the drawn population and on the early path of
prices and displaced demand, and no trust-signal design we tested
reduced either the level or the spread. Isolating true path
dependence, in the sense of \citet{salganik2006}, would require
repeated stochastic histories on a fixed population, which our
design does not include; we flag it as a direct extension. Two
caveats apply. Failure to detect a difference is
not equivalence: with ten runs per condition, level effects smaller
than roughly $0.1$ would go undetected. Reliability is also
payoff-irrelevant here (Section~\ref{sec:model}), so this comparison
of information structures
isolates the coordination role of the trust signal rather than its
value as a predictor of failures. For a
platform, the practical reading is correspondingly narrow: in this
market, neither keeping ratings
live nor replacing them with verified true quality made the
resulting market structure any less variable.

\subsection{Of the four interventions, only capacity disclosure has a
clearly detectable effect}
\label{sec:res-interventions}

\begin{figure*}[!htb]
\centering
\includegraphics[width=\textwidth]{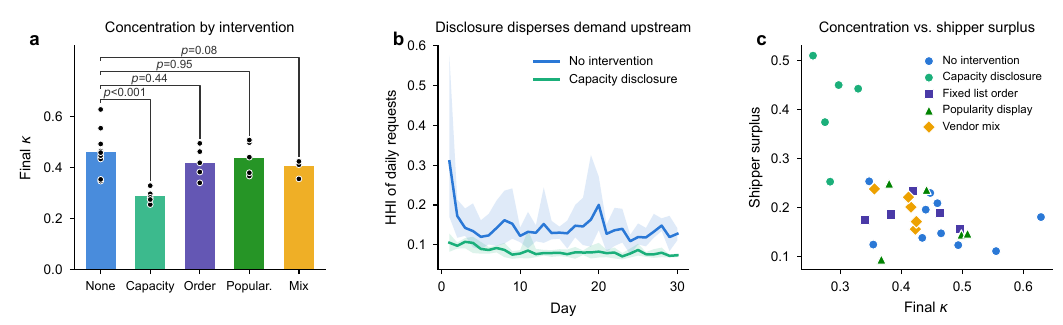}
\caption{\textbf{Capacity disclosure lowers concentration and raises
shipper surplus; the other interventions show no clearly detectable
effect} (GPT, $L=20$).
(a)~Final $\kappa$ by intervention (bars: means; dots: runs).
Brackets give uncorrected two-sided Mann--Whitney $p$-values against
the no-intervention benchmark; only capacity disclosure remains
significant after Holm--Bonferroni correction across the four
comparisons (adjusted $p = 0.003$).
(b)~HHI of daily first choices with and without capacity disclosure
(mean $\pm$ range across runs).
(c)~Final $\kappa$ against shipper surplus, one marker per run.}
\label{fig:interventions}
\end{figure*}

The third lever is what else the platform displays. Starting from the
default condition, we added one intervention at a time and compared final
concentration against the no-intervention benchmark
(Fig.~\ref{fig:interventions}a). Capacity disclosure lowered final
$\kappa$ to $0.29$ (range $0.26$--$0.33$), below every no-intervention
run (Mann--Whitney $p < 0.001$; the only effect that survives
Holm--Bonferroni correction across the four intervention
comparisons, adjusted $p = 0.003$; difference in means $-0.17$,
bootstrap $95\%$ CI $[-0.23, -0.12]$). Fixed list order and popularity
display straddled the benchmark with effects that changed sign from
run to run ($p = 0.44$ and $p = 0.95$; differences in means $-0.04$
and $-0.02$, with CIs $[-0.11, 0.03]$ and $[-0.09, 0.05]$ that
bracket zero); the position bias documented
in evaluation settings \citep{zheng2023} did not survive contact with
price and rating signals. The vendor mix ended slightly below the
benchmark on average (final $\kappa$ of $0.36$--$0.43$ across runs,
mean $0.41$ against $0.46$) without reaching significance
($p = 0.075$; adjusted $p = 0.23$), consistent with the day-one
result that the three
vendors share the same favorite. With five runs the design cannot
distinguish a null from a modest effect, but mixing models that agree
clearly does not diversify demand the way capacity disclosure does.

The mechanism of capacity disclosure operates upstream of the rating
loop (Fig.~\ref{fig:interventions}b); the effect is attributable to
the displayed information itself, not to the sequential execution the
disclosure requires, since prompts carry no information about other
shippers' same-day decisions unless capacity is disclosed. With
remaining capacity on
display, the HHI of daily requests started at $0.11$ on day one,
against $0.31$ without disclosure, so the pile-up that capacity and
prices must undo later never formed. Service failures and their delay
costs all but disappeared, which is where the surplus gain originated:
in the concentration--surplus plane
(Fig.~\ref{fig:interventions}c), capacity disclosure occupies its own
corner, with shipper surplus of $0.41$ against the baseline $0.19$,
while every other intervention clusters with the baseline. The gain
is not a transfer from carriers: total surplus per ton-distance
(shipper value minus carrier resource cost, with unserved loads at
the delay cost, so that payments net out) rose from $0.63$ under the
benchmark to $0.76$ under disclosure, because loads that would have
failed were served.

\subsection{The settled matching: who ends up with whom}
\label{sec:res-matching}

\begin{figure*}[!htb]
\centering
\includegraphics[width=\textwidth]{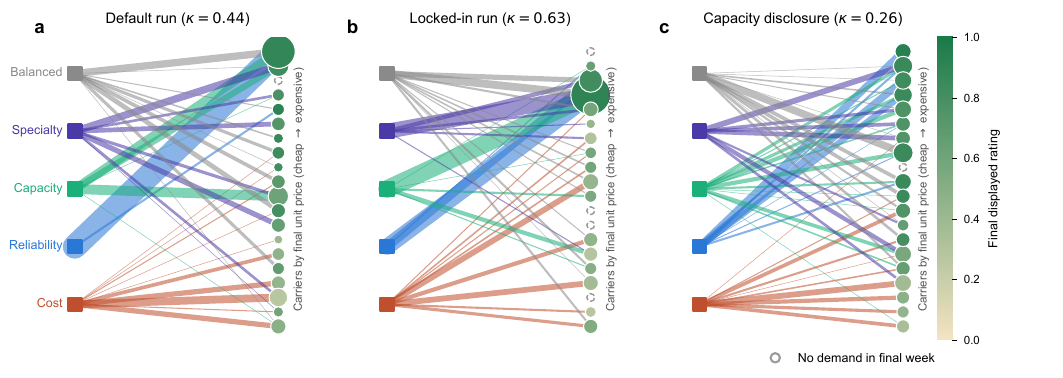}
\caption{\textbf{The settled matching between shipper priorities and
carriers} (first choices, days 26--30).
Left nodes: the five procurement priorities (ten shippers each).
Right nodes: the twenty carriers, ordered bottom to top by final unit
price, sized by first choices received, colored by final displayed
rating; open dashed circles mark carriers that receive no first
choice in the final week. Edge width gives the number of first
choices from a priority to a carrier.
(a)~The default-condition run of Fig.~\ref{fig:dayone}c--d.
(b)~A locked-in run of the same condition.
(c)~A capacity-disclosure run.}
\label{fig:matching}
\end{figure*}

Concentration indices compress the outcome into one number. The audit
trail also shows who ends up matched with whom.
Figure~\ref{fig:matching} draws the final week of first choices (days
26--30) as a network from the five procurement priorities to the
twenty carriers, for three runs that span the results above: the
default-condition run of Fig.~\ref{fig:dayone}c--d, a locked-in run
of the same condition, and a capacity-disclosure run.

Three points emerge. First, the concentration that survives in the
default condition is structured by preference, not arbitrary
(Fig.~\ref{fig:matching}a). Cost-priority shippers spread over the
cheap end of the price range (mean paid unit price $0.50$), while
reliability-priority shippers buy at the expensive top (mean $1.48$),
and the carriers they pick average a true reliability of $0.88$
against $0.45$ for the cost segment; one carrier receives no first
choice at all. The $\kappa = 0.44$ at which this particular run
settles reflects a few favorites
per segment rather than one market-wide favorite. Second, the
locked-in tail of Section~\ref{sec:res-information} is a different
matching structure under identical rules
(Fig.~\ref{fig:matching}b): the favorite absorbs $58$--$68\%$ of the
first choices of the reliability, capacity, and specialty segments
(and none of the cost segment, which keeps buying cheap), while four
carriers end the month without any first choice. Third, capacity
disclosure changes the network, not just the index
(Fig.~\ref{fig:matching}c): demand spreads over nearly every carrier,
the preference sorting at both ends of the price range persists, and
reliability-priority shippers pay $1.09$ instead of $1.48$, because
the pile-up that would have driven their favorites to the price cap
never forms.

\section{Discussion and implications}
\label{sec:discussion}

\subsection{Summary of results}
Our simulations gave a consistent answer to the opening question. When
LLM agents take over carrier selection, their choices concentrate on a
single carrier immediately: the pile-up appeared on day one in every
experiment, at every exposure and under every trust signal, because the
agents share the same preferences before the market has generated any
history. What happened next depended on the market's rules rather than
on the models. The capacity limit spread the excess demand within days
and gave unrated entrants work; congestion pricing set the speed of
that spreading and passed its cost to shippers; and the surviving
concentration varied widely from one sampled market to another under
every
trust signal we tested. Comparing the three levers
by effect size: showing more candidates moved mean concentration from
$0.26$ to $0.43$; replacing disclosed true quality with estimated
ratings moved neither the mean nor the run-to-run dispersion; and
disclosing capacity moved concentration by
$-0.15$ while doubling shipper surplus.

\subsection{Implications for platform design}
Four operational lessons follow for DFM platforms preparing for
algorithmic shippers. First, the length of the candidate list is a
market-structure decision, not a user-interface detail: below roughly
ten displayed candidates the monoculture stays inert, and the
transition zone sits within the range platforms use today. Second,
capacity disclosure is the cheapest effective lever. The platform
already holds the data, the intervention needs one line in the
interface, and in our runs it cut concentration by a third and doubled
shipper surplus by preventing the pile-up rather than repairing it.
Third, interventions that sound protective under-deliver: mixing model
vendors fails because the vendors share the same favorite, and
randomizing or freezing list order washes out against price and rating
signals. Fourth, do not expect better trust signals to stabilize the
market. Neither keeping ratings live nor replacing them with
disclosed true quality reduced the variability of the final market
structure, and most of that variability traced to the drawn market
itself rather than to the signal. The lesson is bounded by the model:
quality here has no payoff consequences, so the signal's only role is
coordination; where quality predicts physical failures, better
signals may still earn their keep through match quality even if they
do not reduce concentration. A platform that wants a predictable
market
structure has to act on the mechanisms that create the lock-in, the
congestion of early demand and its pricing, rather than on the
accuracy of the scores it displays.

The vendor comparison carries its own message. A shipper's choice of
foundation model is market-relevant: Claude populations never
concentrated at any exposure while Gemini populations concentrated
most, so the aggregate market outcome depends on a decision each firm
makes privately for reasons of cost or convenience.

\subsection{Implications for the monoculture debate}
Our experiments speak to both sides of the policy discussion around
algorithmic homogeneity. On one side, they show that static audits of
decision correlation overstate the danger: we found that in a market
with binding capacity and responsive prices, the worst moment is the
first one, after which the market itself undoes part of the
concentration (at the shippers' expense). On the other side, they show
that the correction has limits: concentration settled at $2.4$ times
the random floor, the decline slowed when prices were sticky, and
which carrier ends up dominant varies widely from one sampled market
to another
regardless of the trust signal on display. The practical remedy our experiments single
out is transparency about capacity. This suggests disclosure
obligations as a lighter-touch alternative to regulating the models
themselves, complementing the antitrust concerns raised for LLM
pricing agents \citep{fish2024}. The dispersion result connects to a
pattern now emerging across autonomous-agent operations:
\citet{long2026reliability} document agent bullwhip, run-to-run
decision instability, in serial supply chains, and our market
exhibits its matching-market counterpart, a structure that varies
widely from market to market under every trust signal. In both settings
the instability survives remedies aimed at the signal (repeated
sampling of the model there, better ratings here), which points
attention toward the institutions the agents operate in rather than
toward their outputs.

\subsection{Relation to human freight markets}
The behaviors our agents display have documented human counterparts:
carriers respond to shippers' tendering consistency within a market
cycle \citep{acocella2020}, and
long-term relationships coexist with spot opportunism
\citep{harris2025ltr,harris2025spot}. LLM shippers reproduce the
relationship channel through their transaction history and respond to
spot prices, which makes the comparison meaningful, but we do not
claim behavioral equivalence. A head-to-head benchmark against
estimated human carrier-choice models is the natural next step, and
the audit trail our platform produces makes that comparison feasible
at the level of individual decisions.

\subsection{Limitations}
Five limitations bound the claims, and each points to a specific
extension. First, the model is a stylized capacity-constrained
procurement market inspired by freight rather than a calibrated
freight market: fifty shippers,
twenty carriers, a single stylized set of shipping routes, and
stationary demand over thirty days; contract freight, differences
across routes, and seasonality are absent, origins and destinations
do not affect carrier suitability, specialty is a carrier scalar
rather than a carrier--load match, and distance and tonnage scale
every carrier's quote by the same factor. The economic parameters
were likewise fixed at one point of a larger space (aggregate
capacity at $1.3\times$ expected demand, price cap $1.5$, delay cost
$0.3$, three-tender routing guides, thirty days), and only the
price-adjustment rate received a robustness check; the
capacity-disclosure effect in particular should strengthen as
capacity tightens, and mapping these sensitivities is the immediate
next use of the platform. These features change where
demand piles up, not the fact that correlated preferences pile it up,
so we expect the day-one result to survive them. Two quantitative
questions do depend on scale and deserve dedicated sweeps: whether
the sharp rise in concentration sits at an absolute list length
(about ten
candidates) or at a fraction of the carrier pool (half of twenty),
and how $\kappa$ and the HHI, which mechanically depend on the number
of carriers, translate to larger markets.

Second, carriers price
mechanically and never act strategically: there is no carrier-side AI,
no entry, and no exit. Strategic carriers would likely amplify our
mechanism rather than dampen it, since a carrier that anticipates
LLM preferences can court them, for instance with AI-written pitches
that exploit the documented preference of LLMs for LLM-generated text
\citep{laurito2025}; carrier exit would convert the welfare stakes
from price premia into physical delivery failures
\citep{elliott2022}. Both are worthwhile follow-ups.

Third, quality
is informational rather than payoff-relevant in our market: reviews
are unbiased binary draws, reliability does not cause delivery
failures, and so agents lose nothing by ignoring the trust signal.
This isolates the signal's coordination role, but it also means the
null result of Section~\ref{sec:res-information} should not be read
as ``information quality never matters'': where quality predicts
physical failures, or where reviews inflate and censor as field
studies document \citep{filippas2022,nosko2015}, the comparison could
come out differently. Extensions with a payoff-relevant $\rho$ and
biased review processes are the direct test; the same extensions
would also let quality enter the surplus accounting.

Fourth, results are snapshots of three specific
commercial models at temperature $0.7$ with one English prompt
format. We did not vary prompt wording, and LLM choices are known to
shift with phrasing; the fact that three independently trained models
produced the same day-one favorite suggests the concentration result
is not an artifact of our template, but a systematic prompt-sensitivity
study, together with successor model versions, is needed before the
vendor-specific numbers are treated as stable.

Fifth, and related,
our shipper populations are closer to replicated instances of one
centrally designed agent than to independently engineered procurement
systems: all fifty shippers share the same prompt template, decision
format, temperature, and history structure, differing only in their
stated priority. Real adopters would differ in contracts,
service-level requirements, prompt engineering, tools, and
organizational policy, and even firms using the same foundation model
would implement different agent architectures. The day-one
convergence is therefore best read as an upper bound, the worst case
in which deployments share both the model and the task formulation,
and the vendor-mix condition, which varies only the model, is a weak
test of genuinely independent system design. How much heterogeneity
in agent engineering dissolves the day-one consensus is an open
question the platform can address directly.

\section{Conclusion}
\label{sec:conclusion}
We placed populations of real LLM agents inside a freight matching
market that answers back, with binding capacity, congestion pricing,
and endogenous ratings, and measured what algorithmic procurement does
to market structure. Demand concentrated on a single carrier on the
first day in every condition we tested, so the interesting question is
not whether AI shippers concentrate but what happens next, and that
depends on the market's rules. The capacity limit spread the excess
demand, prices set the speed of the spreading, the surviving market
structure varied widely from market to market under every trust
signal,
and the platform's
information choices, above all how many candidates to show and whether
to reveal remaining capacity, governed the result more than the choice
of model. For transportation research, this relocates the question.
The arrival of AI shippers is not a forecast problem about what the
models will prefer; their preference is uniform and immediate. It is a
design problem about the market they will prefer it in.

\appendix
\renewcommand{\thesection}{Appendix~\Alph{section}}
\setcounter{figure}{0}
\renewcommand{\thefigure}{A\arabic{figure}}

\section{Prompt and example response}
\label{app:prompts}

Each decision is one LLM call. The prompt below is reproduced verbatim
from the audit log (default condition, day 1, one shipper; carrier lines
abbreviated to three of twenty). The system instruction asks for three
ranked carrier identifiers in JSON.

\begin{small}
\begin{verbatim}
You are a decision-making agent in charge of freight
procurement at a shipper (cargo owner) company. From the
list of candidate carriers, choose 3 carriers in order of
preference, identified by their carrier ID (e.g.
"carrier_07"). If your 1st choice is fully booked, your
2nd choice will be tendered, then your 3rd. Respond ONLY
in the following JSON format (no other text before or
after):
{"choices": ["<1st choice>", "<2nd choice>",
 "<3rd choice>"], "reason": "<your rationale in 1-2
 sentences>"}

# Shipper ID: shipper_049
# Your priority: balanced (weigh all factors evenly)
# Today's request: loc_12 -> loc_07, distance 17,
  15 tons, within 1 day
# Summary of your past transactions: none

# Candidate carriers (in random order)
  1) carrier_10: quoted price 138 (unit price 0.54 x
     distance 17 x 15 tons), customer rating 0.42 (based
     on 74 rated shipments), company size: up to 27
     tons/day, specialty 0.65
  2) carrier_05: ...
  3) carrier_01: ...

Choose 3 carriers in order of preference (use their
carrier IDs) and respond in JSON.
\end{verbatim}
\end{small}

The trust-signal line varies by condition: the truth condition shows
\texttt{reliability 0.42}, the static condition shows the same
customer-rating format frozen at its initial value, and the
capacity-disclosure condition appends \texttt{remaining capacity
today: 34 tons} to each line. Carriers without reviews display \texttt{no
customer ratings yet (new carrier)}. The five procurement priorities
are inserted verbatim into the \texttt{Your priority} line:
\texttt{cost first (minimize freight cost)}, \texttt{reliability
first (avoid delivery problems)}, \texttt{capacity first (prefer
carriers that can handle volume)}, \texttt{specialty first (prefer
carriers whose expertise fits the route)}, and \texttt{balanced
(weigh all factors evenly)}; capacity-priority shippers act on the
displayed company size (tons per day), since remaining capacity is
hidden outside the disclosure condition. A typical raw response reads:
\texttt{\{"choices": ["carrier\_18", "carrier\_16", "carrier\_19"],
"reason": "carrier\_18 offers the best overall value with a low quoted
price, a high customer rating, and sufficient capacity for 15
tons."\}}

\section{Reference decision rules}
\label{app:baselines}

\begin{figure*}[!htb]
\centering
\includegraphics[width=\textwidth]{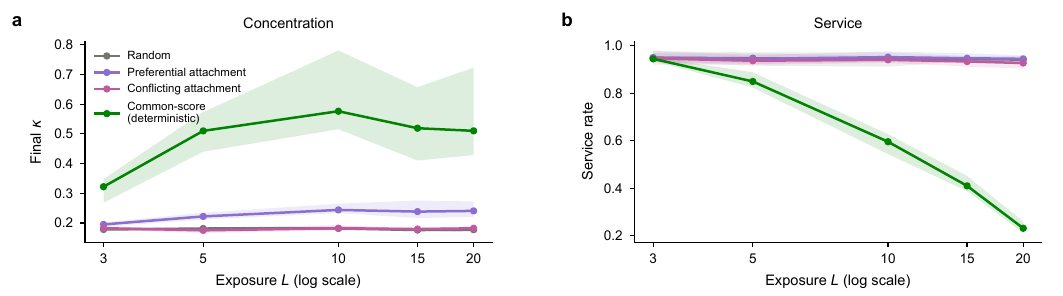}
\caption{\textbf{Reference decision rules across the exposure sweep}
(five runs each; mean $\pm$ range).
(a)~Final $\kappa$. (b)~Service rate.}
\label{fig:baselines}
\end{figure*}

The four reference rules replace the LLM call and receive the same
candidate information. \emph{Random} draws uniformly from the displayed
list. \emph{Preferential attachment} \citep{barabasi1999} draws
carrier $j$ with probability
proportional to $1 + m_j$, where $m_j$ is the carrier's accumulated
matches discounted at $0.95$ per day (the same discounted record that
underlies $\kappa$).
\emph{Conflicting attachment} draws with probability proportional to
$(1 + m_j)\exp(-0.05\,m_j)$, so accumulated matches attract and then
repel; the rule is our construction, and the name borrows from the
conflicting node-attachment mechanism of \citet{leung2016}.
\emph{Common-score} ranks candidates deterministically by one fixed
linear score of the displayed attributes,
$-0.8\,\mathrm{price} + 1.1\,\mathrm{rating} + 0.8\,\mathrm{capacity}
+ 1.0\,\mathrm{specialty}$ (attributes normalized to comparable
scales); it sees the same observed rating the LLM agents see, never
the true reliability (carriers without reviews enter at the
uninformative prior $0.5$, and capacity enters as the normalized
weight underlying the displayed tons per day). Figure~\ref{fig:baselines} reports all four across the
exposure sweep: the two attachment rules stay near the random floor at
every exposure, which is why the main text uses the common-score rule
and the random floor as its contrasts.

\section{Alternative concentration measures}
\label{app:metrics}

\begin{figure*}[!htb]
\centering
\includegraphics[width=\textwidth]{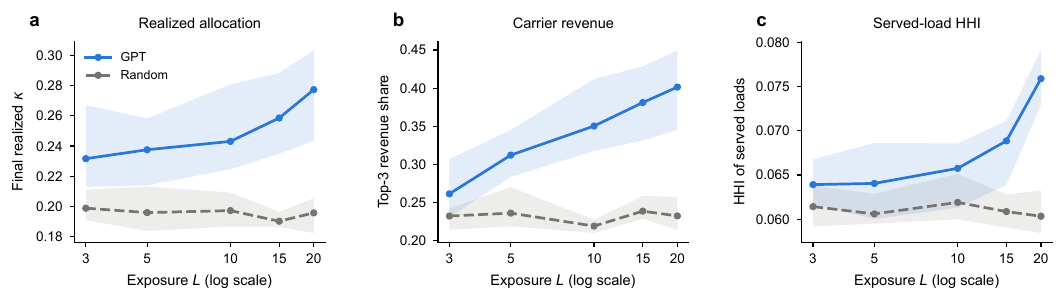}
\caption{\textbf{Alternative concentration measures across the
exposure sweep} (GPT versus random; five runs each; mean $\pm$
range).
(a)~$\kappa$ computed on realized allocations.
(b)~Revenue share of the top three carriers.
(c)~HHI of served loads.}
\label{fig:altmetrics}
\end{figure*}

The main text measures concentration on first choices. Three
alternatives lead to the same conclusions: the realized version of
$\kappa$, computed on the loads carriers end up serving after tender
rejections; the revenue share of the three highest-earning carriers;
and the HHI of served loads. Figure~\ref{fig:altmetrics} traces all
three across the exposure sweep for GPT against the random floor. The
realized measures compress the differences, because the capacity limit
truncates how much demand any carrier can actually serve, but the
ordering and the transition are unchanged. An attempted-tender
version, computed on every tender the waterfall issues rather than on
first choices alone, sits between the two: its final-week top-three
share is $0.39$ (range $0.27$--$0.55$) across the default-condition
runs, against $0.46$ for first choices.

\section*{Declaration of generative AI and AI-assisted technologies
in the writing process}
During the preparation of this work the authors used Claude
(Anthropic) in order to improve the language and readability of the
manuscript and to assist with the implementation of the simulation
and analysis code. After using this tool, the authors reviewed and
edited the content as needed and take full responsibility for the
content of the publication.

\section*{Declaration of competing interest}
The authors declare that they have no known competing financial
interests or personal relationships that could have appeared to
influence the work reported in this paper.

\section*{Data availability}
The simulation platform, prompt templates, experiment scripts, and the
complete decision-level audit logs (every prompt, raw model response,
parsed ranking, and daily market state for all reported cells) are
publicly available at \CodeRepoURL.

\printcredits

\bibliographystyle{cas-model2-names}
\bibliography{refs}

\end{document}